\documentclass[
 reprint,
 secnumarabic,
 amssymb,
 amsmath,
 apl,
 groupedaddress,
 twocolumn,
 preprint,onecolumn,
 frontmatterverbose,
 a4paper
]{revtex4}
\usepackage{graphicx}
\usepackage{bm}
\usepackage{dcolumn}
\usepackage[colorlinks=true,linkcolor=blue,citecolor=blue]{hyperref}%
\newcommand{\textOmega}{$\Omega$}\renewcommand{\textmu}{$\mu$}
\expandafter\ifx\csname package@font\endcsname\relax\else
 \expandafter\expandafter
 \expandafter\usepackage
 \expandafter\expandafter
 \expandafter{\csname package@font\endcsname}%
\fi
\begin{document}
\title{Ionization wave propagation on a micro cavity plasma array}
\author{Alexander Wollny$^1$, Torben Hemke$^1$, Markus Gebhardt$^1$,
Ralf Peter Brinkmann$^1$, Henrik Boettner$^2$, J\"org Winter$^2$,
Volker Schulz-von der Gathen$^2$, Zhongmin Xiong$^3$, Mark J. Kushner$^3$,
and Thomas Mussenbrock$^1$}
\affiliation{$^1$Ruhr University Bochum, Institute of Theoretical
Electrical Engineering, D-44780 Bochum, Germany\\
$^2$Ruhr University Bochum, Institute for Experimental Physics II, 
D-44780 Bochum, Germany\\
$^3$University of Michigan, Department of Electrical Engineering
and Computer Science, 1301 Beal Ave, Ann Arbor, MI 48109, USA}
\date{\today}

\begin{abstract}
Microcavity plasma arrays of inverse pyramidal cavities have been
fabricated in p-Si~wafers. Each cavity acts as a microscopic dielectric
barrier discharge. Operated at atmospheric pressure in argon and
excited with high voltage at about 10~kHz, each cavity develops
a localized microplasma. Experiments have shown a strong interaction
of individual cavities, leading to the propagation of wave-like
optical emission structures along the surface of the array.
This phenomenon is numerically investigated using computer
simulation. The observed ionization wave propagates
with a speed of about 5 km/s, which agrees well the experimental
findings. It is found that the wave propagation is due to sequential
contributions of a drift of electrons followed by drift of ions
between cavities seeded by photoemission of electrons by the plasma
in adjacent cavities.
\end{abstract}

\maketitle

Atmospheric pressure microplasmas have gained increasing attention
since the early work on microhollow cathode discharges by Schoenbach 
et al. \cite{Schoenbach1996}. Although they are highly collisional, 
microplasmas show strong non-equilibrium behavior \cite{Penache2002}. 
Electrons are hot (up to a few eV) while ions and and the neutral gas 
are cold (about room temperature in most cases, though temperatures
of many hundreds Centigrade occur at high power loading).
Microplasmas are characterized by 
high electron densities (up to a few $10^{17}$ cm$^{-3}$) and high
power densities (up to some 100 kW/cm$^3$). Due to these unique
properties microplasmas have found widespread use in technological
and biomedical applications \cite{Becker2006, Morfill2009, Becker2010,
Lee2011}, such as surface treatment, sterilization, and lighting.
At the same time microplasmas continue to show unexpected phenomena
and are therefore scientifically interesting. For example, arrays
of microplasmas having thousands of individual microdischarges have
been fabricated as efficient lighting sources \cite{Eden2005}.
When excited by AC voltages, the initiation of optical emission
from such arrays has shown wave-like phenomena, propagating across
the array at speeds of a few kilometers per second. In this paper,
we discuss results from experimental and computational investigations
of these wave-like propagation of optical emission.

\begin{figure}[t!]
\includegraphics[width=\columnwidth]{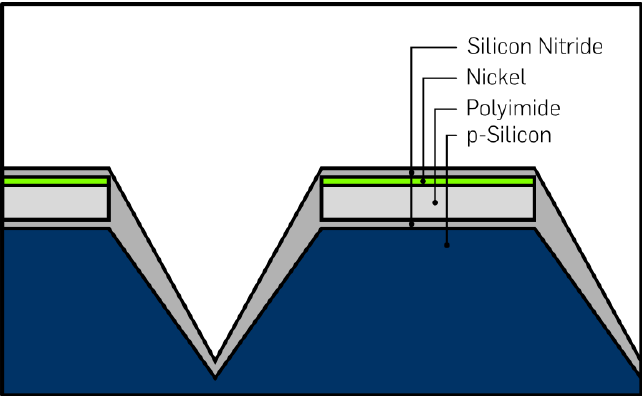}
\caption{Schematics of the modeled micro cavity. p-silicon wafer:
anode, $\phi=0$~V; silicon nitride and polyimide: dielectrics; embedded
nickel grid: cathode, $\phi=-500$~V.\label{fig1}}
\end{figure}

\begin{figure}[t!]
\includegraphics[width=\columnwidth]{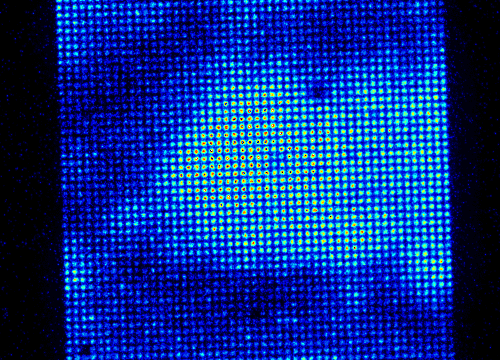}
\caption{Experimental result from phase resolved optical emission
spectroscopy of the microplasma array as registered with an ICCD
camera (false colors) for a gate interval of 100~ns and spectrally
integrated. ($p=10^5$~Pa Ar, $U_{\rm PP}= 830$~V, $f_{\rm ac}=20$~kHz.)
Image taken at $V(t)=-400$~V applied to Ni-grid.\label{fig2}}
\end{figure}

The microcavity array we investigated, developed by Eden et al.
\cite{Eden2005}, has individual inverse pyramidal cavities fabricated
on p-Si wafers (see Fig.~\ref{fig1}), combined with a nickel grid
as a counter electrode and dielectric coatings. Each cavity -- in
essence a microscopic dielectric barrier discharge (DBD) -- has
a base opening of 50 $\times$ 50 \textmu m and a depth of 35
\textmu m. The separation of the individual cavities is 50
\textmu m. The entire array consists of 50 by 50 microplasmas.
When operated with a sawtooth voltage of about 10 kHz in
atmospheric pressure argon, the array emits a bright glow in
the visible wavelength range, which visually appears homogeneous over
the entire array. However, spatially and temporally resolved
emission spectroscopy shows that wave-like structures (see
Fig.~\ref{fig2}) propagate across the array, indicating that
the individual microdischarges strongly interact with each
other \cite{Waskoenig2008, Boettner2010}.

The related computer simulation was performed using the
modeling platform \textit{nonPDPSIM}, described in detail in
\cite{Kushner2004, Babaeva2007, Babaeva2009}. Here we
briefly discuss the relevant physical equations. Poisson’s
equation for the electrostatic potential is self-consistently
coupled with drift-diffusion equations for the charged species
and the surface charge balance equation. The set of equations
is simultaneously integrated in time using an implicit Newton
iteration technique. This integration step is followed by an
implicit update of the electron temperature by solving the
electron energy equation. To capture the non-Maxwellian behavior
of the electrons, the electron transport coefficients and rate
coefficients are obtained by solving the zero-dimensional
Boltzmann's equation for the electron energy distribution.
A Monte Carlo simulation is used to track the trajectories
of sheath accelerated secondary electrons. The transport of
photons is treated by means Green's function method. The
discharge is sustained in argon at atmospheric pressure. The
species in the model are electrons, Ar(3s), Ar(4s), Ar(4p),
Ar$^+$, Ar$^*_2$ and Ar$^+_2$. The photon transport we tracked
in the model is emission dimmer radiation Ar$^*_2$. In the 
absence of impurities and large densities of excited states, this
emission is optically thin. The reaction mechanism is
summarized in Ref. \cite{Bhoj2004a}.

The simulation addresses a sub-system of the micro cavity
array over a time interval of 50~ns. The geometry is
two-dimensional and consists of three neighboring cavities,
which appear in the model as slits. The entire computational
domain is 1~mm tall by 1.5~mm wide, with boundaries of which
are electrically grounded. The conductivity of the wafer in
contact with the boundary is 0.17~\textOmega cm. The secondary
electron emission coefficient by ions is 0.15 on all surfaces.
A photoelectron emission coefficient of 0.05 is assumed on the
dielectric surface. The step function DC potential of $-500$ V
is applied to the embedded nickel electrode grid. 

\begin{figure*}[t!]
\includegraphics[width=\textwidth]{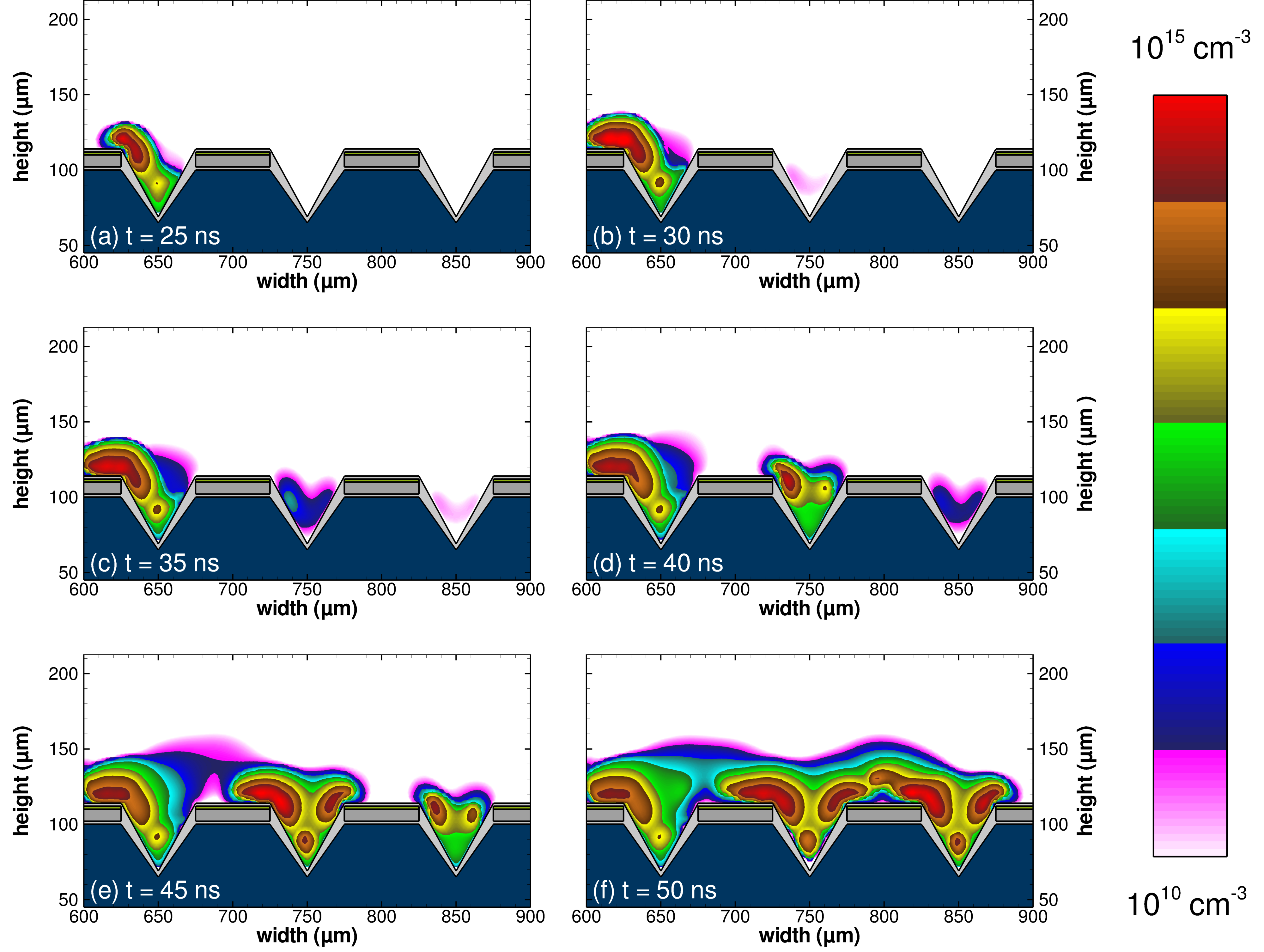}
\caption{Temporal evolution of the electron density distribution in
units of cm$^{-3}$ on logarithmic scale. Figures (a-f) shows the
ignition of the three cavities starting at 25~ns in time steps of
$\Delta t$ = 5~ns. The width and separation of a single cavity is
50 $\mu$m, the overall depth is 45 $\mu$m. The simulation is
performed for Ar as a working gas at atmospheric pressure, a driving
DC voltage of $-500$ V applied to the embedded Ni-grid.\label{fig3}}
\end{figure*}

The initial plasma density is a Gaussian shaped spot of plasma
10 \textmu m wide and $5\times10^{10}$ cm$^{-3}$ in magnitude
at the left vertex of the left cavity. The resulting electron
density as a function of time is shown in Fig.~\ref{fig3}.
Ions accelerated by the negative potential of the nickel grid
drift towards the vertex of the cavity and initiate secondary
electron emission. The secondary electrons are accelerated into
the cavity and generate ions and electrons. The ionization rate
peaks at the vertices at about $5\times10^{24}$ cm$^{-3}$ s$^{-1}$
with an electron temperature of 8 eV, and then spreads on the
top dielectric, similar to a conventional DBD. Meanwhile, photons
produced by relaxation of Ar$_2^*$ lead to photoelectron emission
on the dielectric surfaces, mainly at the right vertex of the
neighboring cavity. At $t=25$~ns a sufficiently large electron
density of about $10^{15}$~cm$^{-3}$ is reached in the left cavity, 
that the potential is shielded from the interior of the left cavity and
the discharge turns from a Townsend mode to a glow mode. The quasineutral
region fills the left cavity and spreads over its edges, charging
the top dielectric. 5~ns after the left cavity is ignited, an
electron density larger than $10^{10}$~cm$^{-3}$ is produced in
the middle cavity initiated by avalanching of photo-generated electrons, 
and another 5~ns later in the right cavity.
Both cavities undergo a transition from a Townsend mode to glow mode
10~ns after the initial electron density is generated
(Fig.~\ref{fig3}d,e). In all three cavities a quasineutral region
develops. These individual plasmas connect after their
expansion out of the cavities, as shown in Fig. \ref{fig3}f. The
electron density peaks in the head of the avalanche with
$n_\mathrm{e}=1.5\times10^{16}$~cm$^{-3}$. Within the cavities
the electron density exceeds $10^{15}$~cm$^{-3}$ leading to a
Debye length $\lambda_\mathrm{D}<1$~\textmu m. During the ignition,
the electron temperature reaches 5~eV within the cavities, but decays
rapidly. The observed densities agree with values expected for this
kind of discharge \cite{Becker2006}.

\begin{figure}[!t]
\includegraphics[width=\columnwidth]{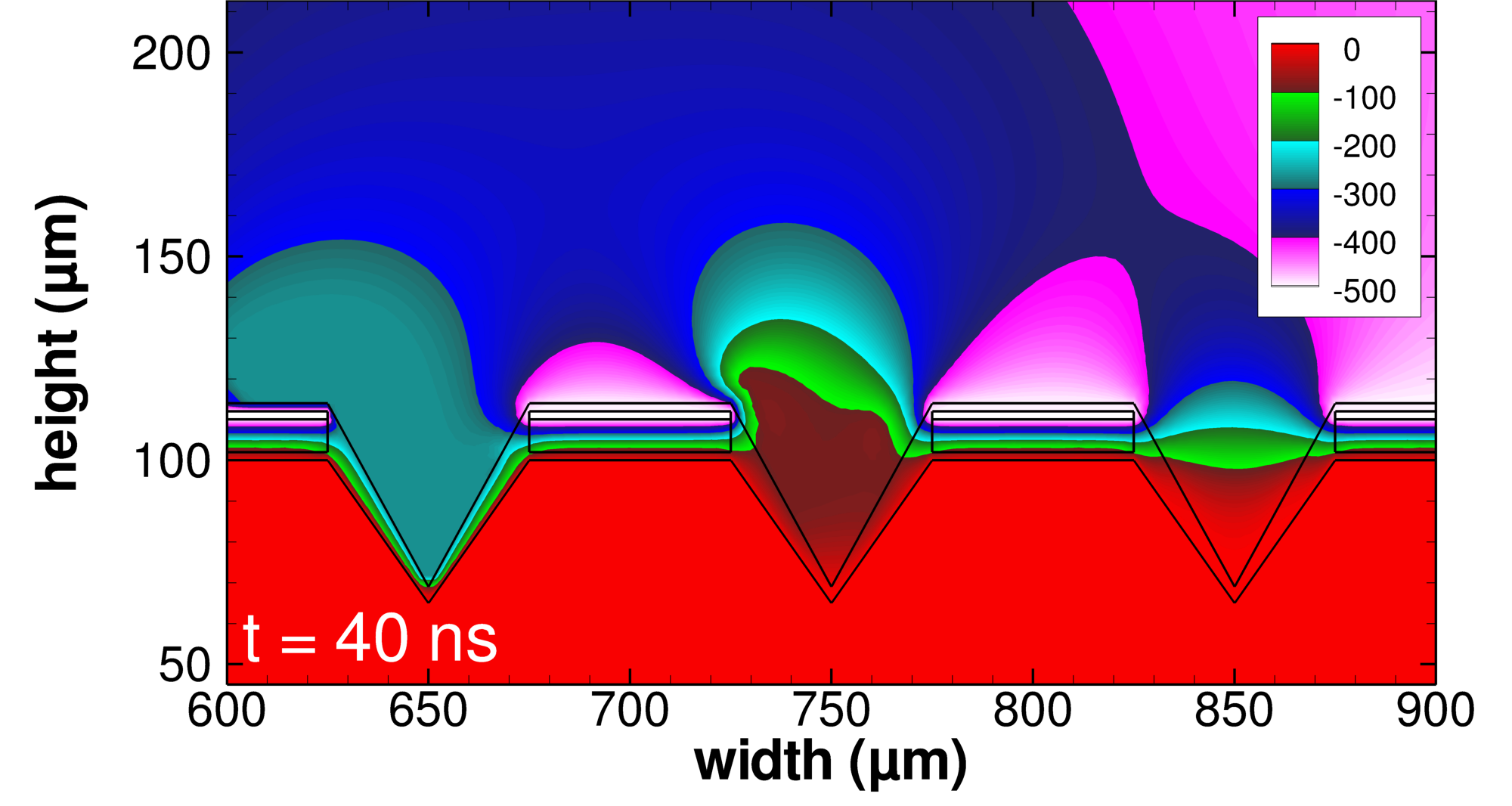}
\caption{Simulated potential at $t = 80$ ns. Left hand cavity is
ignited whereas right hand cavity shows vacuum potential. Centered
cavity is in the stage of transition from Townsend mode to glow
mode.\label{fig4}}
\end{figure}

The electric potential is plotted at $t=40$ ns in Fig.~\ref{fig4}.
It shows the three phases of the discharge development. The potential
in the right cavity is essentially the vacuum potential -- there is
little space charge to influence the electrical field. The left
cavity in contrast contains a quasineutral plasma with the plasma
being nearly equipotential. The majority of the potential is dropped
over the dielectrics which covers the anode and the cathode, as
one expects for a dielectric barrier discharge. The electron
avalanche within the center cavity produces an equipotential channel
which merges with that of the left cavity and indicates the transition
from a Townsend mode to a glow mode discharge.

Taking the time between the ignition of the left and the right cavity
one finds an effective ionization propagation speed of about 5 km/s
which agrees very well with the measured propagation speed of 5 to 10
km/s. In the absence of photoelectron emission, the propagation speed
is about 1 km/s.
Although not all effects are addressed with our simulation,
the ionization wave is well described. Other effects of importance
are surface charges accumulated in a previous half cycle and a second
ionization wave after an further increase of the voltage. The effect
of surface charges accumulated at a previous half cycle with an inverse
voltage results in an enhancement of the electrical field an reduces
the breakdown voltage. Both are a matter of ongoing research. To
summarize, the observed ionization wave is driven by photoelectrons
emitted at neighbored cavities. These electron source causes the ignition
of a single cavity and provides new photons to sustain the ionization
wave propagation.

The authors gratefully acknowledge financial support by the Deutsche
Forschungsgemeinschaft in the frame of Research Group 1123
\textit{Physics of Microplasmas} as well as the \textit{Ruhr University
Research School}.

\clearpage

\clearpage

\clearpage

\clearpage

\end{document}